\documentclass{emulateapj}
\usepackage{apjfonts}

\slugcomment{Draft}
\slugcomment{\today}

\shorttitle{S0 evolution in clusters}
\shortauthors{Poggianti et al.}

\begin{document}


\title{The evolution of spiral, S0 and elliptical galaxies in clusters}


\author{Bianca M. Poggianti$^1$, Giovanni Fasano$^1$, Daniela
Bettoni$^1$, Antonio Cava$^{1,2}$, A. Dressler$^3$, E. Vanzella$^4$, Jesus
Varela$^1$, Warrick J. Couch$^5$, Mauro D'Onofrio$^6$, Jacopo Fritz$^1$, Per
Kjaergaard$^7$, Mariano Moles$^8$, T. Valentinuzzi$^6$}
\affil{$^1$INAF-Astronomical Observatory of Padova, Italy, $^2$Instituto de Astrofisica de Canarias La Laguna, Spain, $^3$The Observatories of the Carnegie institution of Washington, Pasadena, USA, $^4$INAF-Astronomical Observatory of Trieste, Italy $^5$Center for Astrophysics and Supercomputing, Swinburne University of Technology, Australia $^6$Astronomical Department, University of Padova, Italy, $^7$Copenhagen University Observatory. The Niels Bohr Institute for Astronomy, Physics and Geophysics, Denmark, $^8$Instituto de Astrofisica de Andalucia, CSIC, Granada, Spain}





\begin{abstract}
We quantify the evolution of the spiral, S0 and elliptical fractions
in galaxy clusters as a function of cluster velocity dispersion
($\sigma$) and X-ray luminosity ($L_X$) using a new database of 72
nearby clusters from the WIde-Field Nearby Galaxy-cluster Survey
(WINGS) combined with literature data at $z=0.5-1.2$. Most WINGS
clusters have $\sigma$ between 500 and 1100 $\rm km \, s^{-1}$, and
$L_X$ between 0.2 and $5 \times 10^{44} \rm erg/s$.  The S0 fraction
in clusters is known to increase with time at the expense of the
spiral population. We find that the spiral and S0 fractions have
evolved more strongly in lower $\sigma$, less massive clusters, while
we confirm that the
proportion of ellipticals has remained unchanged. 
Our results demonstrate that morphological
evolution since $z=1$ is not confined to massive clusters, but is actually more
pronounced in low mass clusters, and therefore must originate either
from secular (intrinsic) evolution and/or from environmental
mechanisms that act preferentially in low-mass environments, or both
in low- and high-mass systems. We also find that the evolution of the
spiral fraction perfectly mirrors the evolution of the fraction of
star-forming galaxies.
Interestingly, at low-z the spiral
fraction anticorrelates
with $L_X$. Conversely, no
correlation is observed with $\sigma$. 
Given that both $\sigma$ and $L_X$ are 
tracers of the cluster mass, these results pose a challenge for
current scenarios of morphological evolution in clusters.
\end{abstract}





\section{Introduction}

The strongest evidence that a significant fraction of today's
early-type galaxies have evolved from later-type galaxies at
relatively recent cosmic epochs comes from galaxy cluster studies. HST
images revealed that spirals are proportionally much more common, and
S0 galaxies much rarer, in distant than in nearby clusters, suggesting
that many of the local cluster S0s have evolved from spirals (Dressler et
al. 1997, Fasano et al. 2000, Treu et al. 2003, Postman et al. 2005,
Smith et al. 2005, Desai et al. 2007, but see Andreon 1998
and Holden et al. 2009 for an opposite view).  While the paucity of the S0
population has been observed out to $z \sim 1$, it seems that most of
the morphological evolution occurs at $z \leq 0.4$, because the S0 and
spiral galaxy fractions flatten out in clusters at $z>0.45$ (Desai et
al. 2007). Interestingly, no significant evolution has been detected
in the proportion of cluster ellipticals since $z\sim 1$ (see
references above).

The latest high-z studies of galaxy morphologies in clusters have
started to assess how the morphological mix depends on the global
cluster properties, in particular on $L_X$ and $\sigma$, both considered
rather good tracers of the cluster mass (Postman et al. 2005, Desai et
al. 2007).  Early studies at $z\sim 0$ had found that the
proportions of spirals and early-type galaxies vary with $L_X$ 
(Bahcall 1977, McHardy 1978, Edge \&
Stewart 1991).  Clearly, it is necessary to quantify the morphological
evolution as a function of cluster mass, in order to perform a
meaningful comparison between high- and low-z data and to investigate
the origin of the morphological variations with global environment.

To date, a quantitative analysis of the morphological evolution for
clusters of different masses has been impossible, due to the lack of a
local large sample of clusters with similarly detailed informations as
the high-z samples.  In this Letter, we present the comparison between
a new dataset of 72 clusters in the local universe and the distant
clusters currently available in the literature, quantifying the
evolution of the fractions of spiral, S0 and elliptical galaxies in
clusters with a wide range of velocity dispersions and $L_X$.


Throughout this paper we use ($H_0$, ${\Omega}_m$, ${\Omega}_{\lambda}$) = 
(70,0.3,0.7).

\section{Observations}

The WIde-Field Nearby Galaxy-cluster Survey (WINGS) is a
multiwavelength photometric and spectroscopic survey of 77 galaxy
clusters at $0.04<z<0.07$ (Fasano et al. 2006).  Clusters were
selected in the X-ray from the ROSAT Brightest Cluster Sample and its
extension (Ebeling et al. 1998, 2000) and the X-ray Brightest
Abell-type Cluster sample (Ebeling et al. 1996).\footnote{X-ray 
luminosities (0.1-2.4keV) from Ebeling et al. (1996,
1998, 2000) have been converted to the cosmology used in this paper.}  
WINGS clusters cover a wide range of $\sigma$, typically
between 500 and 1100 $\rm km \, s^{-1}$, and $L_X$, typically $0.2-5
\times 10^{44} \rm erg/s$.

WINGS has obtained wide-field optical photometry ($BV$) 
for all 77 fields (Fasano et
al. 2006, Varela et al. 2009), as well as infrared
($JK$) photometry (Valentinuzzi et al. 2009), optical spectroscopy
(Cava et al. 2009), U band and $\rm H\alpha$ narrow-band imaging for a
subset of the WINGS clusters.  In this Letter we consider 72 WINGS
clusters, having excluded only clusters with no reliable velocity
dispersion measurement (A311, A2665, A3164 and Z1261) or insufficient
$V$-band seeing quality (A3562).  Hereafter, we use galaxy $V$ total
magnitudes (Sextractor MAGAUTO) from Varela et al. (2009).  
Cluster velocity dispersions were computed combining WINGS and
literature redshifts. For all but one cluster, they are based on more
than 20 spectroscopic members, with an average of 92 spectroscopic
members per cluster (Cava et al. 2009).

Our analysis is based on galaxy morphological types, derived from $V$
images using the purposely devised tool MORPHOT (Fasano et al. 2009,
in preparation).  Our approach is a generalization of the
non-parametric method proposed by Conselice et al. (2000), see also
Conselice (2003). In particular, we have extended the classical CAS
(Concentration/Asymmetry/clumpinesS) parameter set by introducing a
number of additional, suitably devised morphological indicators, using
a final set of 10 parameters. A control sample of $\sim 1000$ visually
classified galaxies has been used to calibrate the whole set of
morphological indicators, with the aim of identifying the best sub-set
among them, as well as of analyzing how they depend on galaxy size,
flattening and S/N ratio.

The morphological indicators 
have been combined with two independent methods, a Maximum Likelihood
analysis and a Neural Network trained on the control sample
of visually classified galaxies. The final, automatic morphological
classification combines the results of both methods. It has been 
carefully verified that our morphological 
classifications reproduce visual classifications by two of us
with a robustness
and a reliability 
comparable with those obtained from visual inspection by experienced
independent classifiers (Fasano et al. 2009, in prep.).

In the following, we use three broad morphological classes,
ellipticals, S0s and ``spirals'', where the spiral class includes
any galaxy later than an S0.  

As high-z comparison, we use the morphological fractions of the ESO
Distant Cluster Survey (EDisCS) sample of 10 clusters at $z=0.5-0.8$
from Desai et al. (2007), and the 5 clusters at $z=0.8-1.2$ from
Postman et al. (2005) for which a $\sigma$ is available.\footnote{For the
Cl 1604+4304 and Cl1604+4321 clusters we have used revised 
1.0 $h_{70}^{-1} \rm \, Mpc $ velocity dispersions
from Gal et al. 2008.}  The
morphological classification of galaxies in these clusters is based on
visual classification of HST/ACS F814W images sampling the rest-frame
$\sim 4300-5500\AA$ range. The reliability of the high-z morphologies
have been tested and discussed at length by Postman et al. (2005) and
Desai et al. (2007), and by previous works.  For the high-z clusters
we use the X-ray luminosities in the 0.5-2keV band provided by Olivia
Johnson (2008 private communication) and from Johnson et al. (2006)
and Postman et al. (2005).

The analysis for both WINGS and high-z clusters includes only galaxies
within $0.6R_{200}$\footnote{$R_{200}$ is defined as the radius
delimiting a sphere with interior mean density 200 times the critical
density, approximately equal to the cluster virial
radius. $0.6R_{200}$ roughly corresponds to $R_{500}$, whose interior
mean density is 500 times the critical density.  The variation of the
morphological fractions between 0.6 and 1$R_{200}$ in Postman et
al. (2005) is negligible (their Table~3). A few WINGS clusters have
photometric coverage slightly smaller than $0.6R_{200}$ (Cava et
al. 2009). None of the results in this paper change if these
clusters are excluded from the analysis.}  brighter than $M_V=-19.5$
for WINGS\footnote{63\% of the galaxies fulfilling these criteria are
spectroscopically confirmed members.} 
and $M_V=-20$ for EDisCS.  The different magnitude limit
accounts for passive evolution dimming between the mean WINGS redshift
and $z\sim 0.6$.  The fractions taken from Postman et al. (2005)
already take passive evolution into account, corresponding to
$M_V=-19.3$ at $z=0$. High-z morphological fractions are corrected for
fore-background contamination by the authors. For WINGS clusters, we
have verified that the field contamination is entirely negligible
using the local field counts and morphological mix by Whitmore et
al. (1993).

This is the first time the morphological evolution in clusters can be
investigated as a function of a proxy for the cluster mass, such as
velocity dispersion or X-ray luminosity. This is possible thanks to
the combination of the first high-z (EDisCS) and first low-z (WINGS)
homogeneous samples with simultaneously a large range in cluster
masses, detailed galaxy morphologies and accurate velocity
dispersions.

\section{Results: the evolution with redshift of the morphological mix}
Figure~1 shows the fractions of spiral, S0, elliptical and early-type
(E+S0) galaxies as a function of cluster velocity dispersion.

At low-z (black points), none of the morphological fractions show a
trend with velocity dispersion.  Spearman probabilities of a
correlation are lower than 70\% in all cases, and least square fits
yield a practically flat best fit. On average, spirals are
23\%$\pm9$\% of all galaxies, S0 galaxies represent almost half of
the cluster population (44\%$\pm10$\%) and ellipticals are about a third
with an average fraction of 33\%$\pm$7. The quoted errors on
these average values correspond to the rms, and represent the
cluster-to-cluster scatter in the morphological fractions at a given
$\sigma$.\footnote{We note that the most massive clusters with $\sigma
\geq 1000 \, \rm km \, s^{-1}$ possibly display a lower scatter than
lower $\sigma$ systems, and tend to have among the lowest spiral
fractions, and highest eraly-type fractions, but the number of very
massive clusters is too low to draw robust conclusions.} For all
morphological types, this cluster-to-cluster scatter is comparable or
even smaller than the average of all poissonian errorbars on the
morphological fractions of individual clusters (Fig.~1), indicating a
quite remarkable homogeneity of the morphological mix in nearby
clusters, regardless of velocity dispersion, over a large range of
$\sigma$.

In contrast, mild correlations both with $\sigma$ and $L_X$ have been
reported in distant clusters by Postman et al. (2005) and Desai et
al. (2007) (see also Figs.~1 and ~2).  Comparing with clusters at
$z=0.5-1.2$ (red points in Fig.~1), we find that {\it the
morphological evolution has been much stronger in progressively lower
velocity dispersion systems}.  As shown in Fig.~1 and summarized in
Table~1, on average, the spiral fraction has evolved from $\sim$60\%
for systems with $\sigma = 500 - 700 \, \rm km \, s^{-1}$, and from
$\sim$40\% for more massive clusters, to $\sim$23\% at low-z
regardless of velocity dispersion.\footnote{ Since the morphological
fractions are flat with $\sigma$ at low redshift, the evolution of the
cluster $\sigma$'s with redshift is irrelevant when assessing the
morphological evolution.}
Similarly, the average S0 fraction has changed from $\sim$15\%
for the least massive clusters, and from $\sim$22\%
for the most massive clusters, to $44$\% in nearby clusters of all $\sigma$'s.
Finally, the evolution of the average elliptical fraction for massive clusters 
is negligible,
as noted by previous
works (eg. Dressler et al. 1997). 

\begin{table}
\begin{center}
\caption{Average morphological fractions.}
\begin{tabular}{lccc}
\tableline\tableline
 & \multicolumn{2}{c}{High-z}  &  {\bf Low-z} \\
\tableline
 & Low-$\sigma$ & High-$\sigma$ & All-$\sigma$ \\
\tableline\tableline
Spirals & 60 & 40 & {\bf 23} \\
S0s     & 15 & 22 & {\bf 44} \\
Ellipticals & 25 & 37 & {\bf 33} \\
\tableline
\end{tabular}
\tablecomments{Low-$\sigma$ and high-$\sigma$ clusters are those
with $\sigma = 500-700 \rm \, km \, s^{-1}$ and $>700 \rm \, km \,
s^{-1}$, respectively. At low redshift, the average fractions are
constant with velocity dispersion (see text).}  
\end{center}
\end{table}


\subsection{Comparison with the evolution of the star-forming fraction}

We now compare the evolution of the morphological fractions with the
evolution of the fraction of star-forming galaxies.  The star-forming
fraction as a function of cluster $\sigma$ was derived as the
proportion of [OII]-line emitters, defined to have an [OII] equivalent
width $\geq 3$\AA, for the EDisCS sample and for an SDSS nearby
cluster sample by Poggianti et al. (2006), for the same galaxy
magnitude limits used here and within $R_{200}$. The best fit trends
of star-forming fractions at high- and low-z (blue lines) are
reproduced and compared to the spiral fractions in the leftmost panel
of Fig.~1 (see also the rightmost panel for the complementary
passively evolving (=non-emission-line) versus early type comparison).

The qualitative and, roughly speaking, also quantitative
correspondence between spiral and star-forming fractions (and,
equivalently, early-type and passively evolving fractions) is striking
both for nearby and distant clusters.  It is well known that not all early-type
galaxies are passively evolving and not all spirals are star-forming,
as testified by the (small) number of ellipticals and S0s with
emission lines and the (significant) number of passive spirals
observed in cluster spectroscopic surveys (eg. Dressler et al. 1999,
Sanchez-Blazquez et al. 2009). However, the agreement in Fig.~1
indicates that, overall, the morphological evolution and the evolution
in the proportion of star-forming galaxies have been very similar. In
spite of the evidence that morphological transformations occur on a
longer timescale than star formation changes (Poggianti et al. 1999),
this agreement suggests that a common physical cause is likely to
regulate both on the long run. In this respect, it is interesting to note
that ``morphology'' is more strongly affected by star formation, and depends
more strongly on environment, than ``galaxy structure'' as measured
for example by the Sersic index, that mainly depends on galaxy mass
(van der Wel 2008).

\subsection{Trends with X-ray luminosity}

Finally, we analyze the morphological trends as a function of $L_X$ (Fig.~2).
The morphological evolution is similar to that
observed for $\sigma$: at a given $L_X$, distant clusters have higher
spiral fractions, and lower S0 fractions, than local
clusters. Elliptical fractions 
are comparable at high- and low-z.

Interestingly, in contrast with the lack of correlation with $\sigma$, 
and in spite of the fact that
both $\sigma$ and $L_X$ are considered tracers of cluster mass, we
find that the 
spiral fraction is statistically
anticorrelated with $L_X$, with a Spearman probability of 99.6\% (Fig.~2). 
The best fit linear relation and its error computed using bootstrap
resampling is $f_{sp} = -0.077\pm0.026 \, log \, L_X + 3.62\pm1.15$.

This confirms previous results that found similar trends between
morphological fractions and $L_X$ (Bahcall 1977, McHardy 1978, Edge \&
Stewart 1991). We note that our correlations are less steep with $L_X$
than in these previous works, possibly due to the fact that we limit
our analysis to a fixed fraction of $R_{200}$, while previous studies
included all galaxies within a plate field-of-view.

\section{Discussion and Conclusions}

The results presented in this Letter highlight two main issues that
are central to understanding why galaxy morphologies change.

First, our findings demonstrate that morphological evolution does not
occur exclusively in massive clusters, being actually more conspicuous
in low mass clusters. This can be due either to ``secular'' galaxy
evolution (via processes intrinsic to each galaxy), or to one or more
environmental processes that work preferentially in low-mass systems,
or both in low- and high-mass systems.

Finding correlations of galaxy properties with global parameters like
cluster mass, velocity dispersion or X-ray luminosity does not
necessarily imply that environment at the epoch of observation is the
agent of galaxy evolution.  The principal cause could be `secular,' by
which we mean that the evolution of morphology or star formation rests
on some fundamental galaxy property/ies, such as galaxy mass, or
galaxy matter density at such an early epoch as to be considered
``initial conditions.''  For example, the epoch at which star
formation will end and the galaxy's morphology evolve --- the galaxy's
internal 'clock' --- could be driven by the primordial density and
mass of available gas, conditions that also correlate with the
properties of the future global environment, such as the cluster
$\sigma$ that we have observed.  
In other words, initial conditions (the ``primordial environment'')
would predetermine individual galaxy properties, in a statistical
sense. 
Trends with global environments as those
we observe would arise because the {\it distribution} of initial conditions
varies systematically with global environmental properties, such as 
cluster mass.

If, instead, environmental effects are directly inducing the
morphological change close to the time we observe it, our results
show that the
processes must work effectively either in both low-mass and high-mass
clusters, or preferentially in the low-mass cluster range of our
sample, and/or perhaps also in groups with even lower masses than those
considered here.
In the case of environmental effects, 
the evolution of morphological fractions
depends on $\sigma$ because the average cluster growth history
(the history of accretion of other clusters, groups and isolated
galaxies onto a cluster) varies systematically 
with cluster mass (Poggianti et
al. 2006).


Clearly, the constancy of the spiral, S0 and elliptical fractions with
$\sigma$ observed in WINGS at the present epoch cannot continue to
systems of even lower velocity dispersions all the way down to
isolated galaxies, otherwise the morphological mix of the general
field at low-z should be the same as in clusters, whereas it is
markedly different.  Somewhere below $\sigma = 500 \rm \, km \, s^{-1}$ 
the average
fraction of spirals must break or slowly transition to the high
fraction of spirals found in the low-density ``field"
population. Understanding the minimum cluster/group mass at which
morphological transformations regularly take place is of course
fundamental not only to identifying the relevant mechanisms at work, 
but also to assessing how many galaxies can be affected and how
this fits in with the general cosmic evolution from
star-forming/late-type to passive/early-type galaxies; for example,
van der Wel. et al. 2007 find no significant evolution in the
early-type galaxy fraction among field galaxies.

In this respect, it is important to note that the limit of the WINGS
survey, $\sigma \sim 500 \rm \, km \, s^{-1}$, corresponds to the
observed break in the trend of star-forming fraction versus $\sigma$
in SDSS clusters at $z=0$: above $500 \rm \, km \, s^{-1}$
the fraction is flat with
$\sigma$ (blue line in Fig.~1), while below the scatter becomes
very large, with the average fraction possibly increasing towards
lower velocity dispersions (see Figs.~4 and ~6 in Poggianti et
al. 2006).  It is reasonable to expect that the morphological
fractions in low $\sigma$ systems continue to follow the trends of the
star-forming fraction (as they do above $500 \rm \, km \, s^{-1}$),
and therefore
to expect a large scatter and an increasing average spiral
fraction beginning just below $500 \rm \, km \, s^{-1}$. If this
is the case, and if the evolution is not secularly, but
environmentally, driven, then some process is probably
affecting galaxies in some, but apparently not all, groups.  This is
consistent with the fact that, at all redshifts, groups are a very
heterogeneous population, some having galaxy properties
resembling massive clusters --- low spiral and star-forming
fractions, --- with others resembling the field (P06, Jeltema et al. 2007,
Wilman et al. 2005, Poggianti et al. 2008, Kautsch et al. 2008, Wilman
et al. 2009, Poggianti et al. 2009).

In conclusion, our result suggests that the candidate environmental
mechanism(s) should be one (those) that act efficiently and quite
universally in clusters with $\sigma \geq 500 \rm \, km \, s^{-1}$,
and probably in {\it some}, but not all, lower-mass clusters and
groups.\footnote{The hypothesis that
late-type galaxies suffer a significant morphological change in {\it
all} systems with masses much lower than $500 \rm \, km \, s^{-1} \sim
10^{14} \, M_{\odot}$ can be ruled out by the fact that the
early-type/passive fraction in local clusters should then be much
lower than observed (Fig.~16 in Poggianti et al. 2006).}
Neither tidal effects, nor processes related to the
intra-cluster medium, can be easily ruled out at this stage (see
Poggianti et al. (2009) and Boselli \& Gavazzi (2006) for a discussion
of the possible mechanisms).

The second challenge for any acceptable evolutionary scenario is posed
by the existence and lack, respectively, of a dependence of the
spiral fraction on $L_X$ and $\sigma$ in local clusters, described
above.  $L_X$ and $\sigma$ are rather well correlated for all local
cluster samples including WINGS (Cava et al. 2009), and both are
considered rather good tracers of cluster mass, albeit with caveats.
It is therefore surprising that the morphological distribution 
correlates at the 3$\sigma$ level with $L_X$, but not with $\sigma$.
Whether this
is due to the way $L_X$ and $\sigma$ differently trace the cluster
mass, whether it is pointing to some residual trend with the
properties of the intra-cluster medium ($L_X$) on top of a more
general evolution, or whether this is support for a secular mechanism
where the correlation of properties does not imply a causal
connection, is an open question.

\acknowledgments 
We would like to thank the anonymous referee for the constructive
and thought-stimulating report.  We warmly thank Olivia Johnson
for kindly computing the 0.5-2keV EDisCS luminosities for us, and
Vandana Desai for sending the high-z fractions in electronic form. BMP
acknowledges the Alexander von Humboldt Foundation and the Max Planck
Instituut fur Extraterrestrische Physik in Garching for a very
pleasant and productive staying during which the work presented in
this paper was carried out. BMP acknowledges financial support from
the FIRB scheme of the Italian Ministry of Education, University and
Research (RBAU018Y7E) and from the INAF-National Institute for
Astrophysics through its PRIN-INAF2006 scheme.



{\it Facilities:} \facility{INT(WFC), 2.2m(WFC), AAT(2dF), WHT(WYFFOS)}.

\begin{figure*}
\vspace{-10cm}
\centerline{\hspace{-0.5cm}\includegraphics[width=0.9\columnwidth]{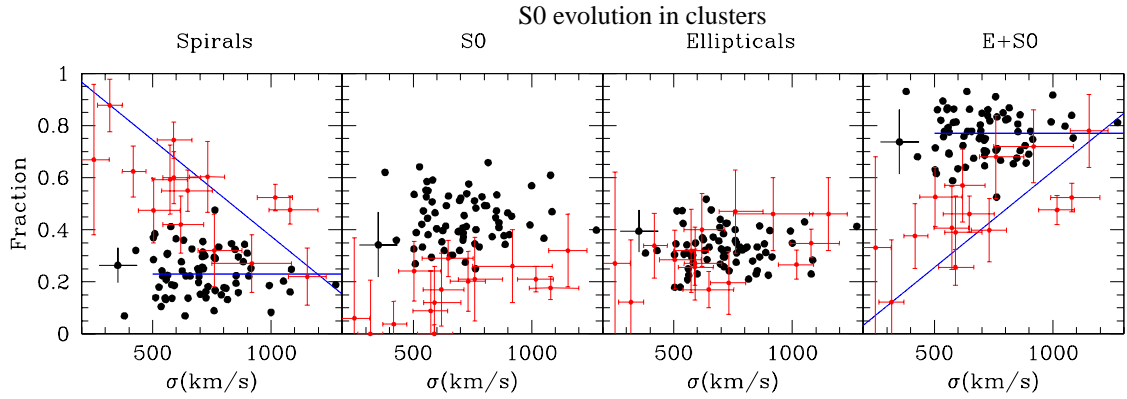}\hfill}
\caption{Morphological fractions within $0.6R_{200}$ versus rest frame
cluster velocity dispersion for galaxies with a passively evolving
absolute magnitude limit $M_V<-19.5$ at the WINGS redshift.  Nearby clusters
(WINGS) are black circles.  Average Poissonian errorbars for WINGS are
shown in each panel on the leftmost datapoint.  Distant clusters
(EDisCS+Postman) are red circles with individual errorbars. None of  
the WINGS fractions is statistically correlated with $\sigma$.  
Solid blue lines in the leftmost panel represent
the average fractions of emission-line galaxies at high-z (inclined
line, EDisCS) and low-z (horizontal line, SDSS cluster sample) 
from Poggianti et al. (2006). Similarly, the solid blue lines 
in the rightmost panel are
the average fractions of passively evolving (non-emission-line)
galaxies.
 \label{letter}}
 \end{figure*}

 \begin{figure*}
\vspace{-10cm}
\centerline{\hspace{-0.5cm}\includegraphics[width=0.9\columnwidth]{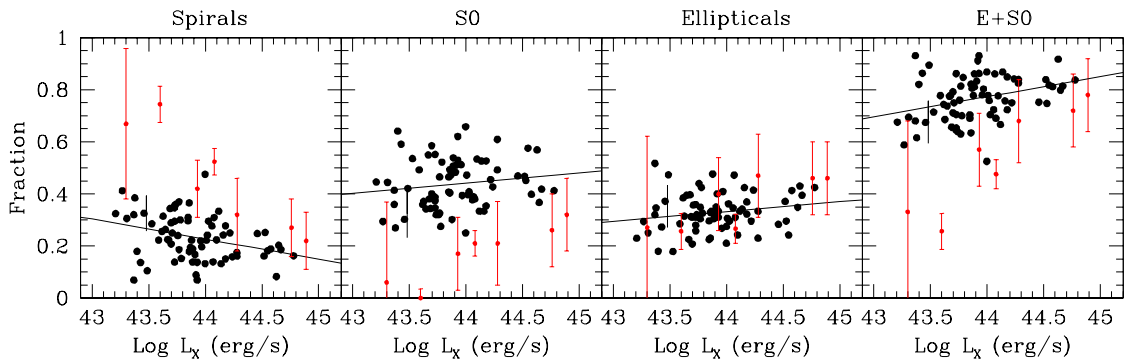}\hfill}
\caption{Morphological fractions versus cluster X-ray
luminosity. Nearby clusters (WINGS) are black circles, with errors on
fractions as in Fig.~1. Distant clusters (EDisCS+Postman with
available $L_X$) are red circles. The WINGS spiral fraction is
statistically anticorrelated with $L_X$ (99.6\%). 
The least square fit is shown in each panel as a solid line.
 \label{letter2}}
 \end{figure*}

\end{document}